\documentclass[12pt]{article}
\usepackage{amsmath}
\usepackage{cite}
\usepackage{graphicx}
\usepackage{amsfonts}
\usepackage{amssymb}
\usepackage{appendix}
\providecommand{\U}[1]{\protect\rule{.1in}{.1in}}

\csname @addtoreset\endcsname{equation}{section}
\newcommand{\eq}{\begin{equation}}
\newcommand{\feq}{\end{equation}}
\newcommand{\eqn}{\begin{eqnarray}}
\newcommand{\feqn}{\end{eqnarray}}
\newcommand{\arr}{\begin{eqnarray*}}
\newcommand{\farr}{\end{eqnarray*}}

\newcommand{\bea}{\begin{eqnarray}}
\newcommand{\eea}{\end{eqnarray}}

\begin{document}
\begin{titlepage}
\begin{center}
\baselineskip=16pt
{\Large\bf One-variable Metrics in String Theory}
\vskip 0.15in
\vskip 10.mm
M. Fakhredin and W. A. Sabra
\vskip 1cm
{\small{\it
Department of Physics, American University of Beirut, Lebanon }}
\vspace{6pt}
\end{center}
\vskip 0.2in
\par
\begin{center}
{\bf Abstract}
 \end{center}
\begin{quote}
Exact solutions depending on one variable of gravitational theory with antisymmetric tensor and a coupled dilaton field are obtained in arbitrary space-time dimensions. These solutions are relevant to M-theory, type IIA and type IIB supergravity theories 
in various space-time signatures. 
\vskip 2.mm
\end{quote}
\hfill
\end{titlepage}

\section{INTRODUCTION}
A notable vacuum solution of four-dimensional Einstein gravity is the
Bianchi I metric \cite{macel}, normally referred to in the literature as the
Kasner metric \cite{kasner}. This metric can be represented by 
\begin{equation}
ds^{2}=-d\tau ^{2}+\tau ^{2p_{1}}dx^{2}+\tau ^{2p_{2}}dy^{2}+\tau
^{2p_{3}}dz^{2}\text{ },  \label{bian}
\end{equation}%
where the so-called Kasner exponents $p_{1}$, $p_{2}$ and $p_{3\text{ }}$are
constants satisfying the conditions 
\begin{equation}
p_{1}+p_{2}+p_{3}=p_{1}^{2}+p_{2}^{2}+p_{3}^{2}=1\text{ }.
\end{equation}%
In $D$-dimensional gravitational theory with arbitrary signature, Kasner
metric is given by \cite{harvey} 
\begin{equation}
ds^{2}=\epsilon _{0}d\tau ^{2}+\epsilon _{\mu }\tau ^{2p_{\mu }}\left(
dx^{\mu }\right) ^{2}
\end{equation}%
with the exponents $p_{\mu }$ satisfying the conditions 
\begin{equation}
\sum_{\mu =1}^{D-1}p_{\mu }=\sum_{\mu =1}^{D-1}p_{\mu }^{2}=1
\end{equation}%
and $\epsilon _{0}$ and $\epsilon _{\mu }$ are constants taking the values $
\pm 1.$

The original metric of \cite{kasner} when generalized to arbitrary
space-time signature \cite{harvey} takes the form 
\begin{equation}
ds^{2}=\epsilon _{0}x^{2a_{1}}dx^{2}+\epsilon _{1}x^{2a_{2}}dy^{2}+\epsilon
_{2}x^{2a_{3}}dz^{2}+\epsilon _{3}x^{2a_{4}}dw^{2}  \label{ok}
\end{equation}%
with the conditions 
\begin{equation}
a_{2}+a_{3}+a_{4}=a_{1}+1,\text{ \ \ \ }a_{2}^{2}+a_{3}^{2}+a_{4}^{2}=\left(
a_{1}+1\right) ^{2}.
\end{equation}%
The cosmological synchronous Bianchi I metric (\ref{bian}) can be obtained
from (\ref{ok}) through a change of variable or simply by setting $a_{1}=0$.
In the analysis of \cite{landau} of the vacuum solutions depending on one
variable, one starts with the metric 
\begin{equation}
ds^{2}=\epsilon _{0}d\tau ^{2}+g_{ij}(\tau )dx^{i}dx^{j},  \label{mek}
\end{equation}%
and obtains, in addition to the Bianchi I metric, non-diagonal vacuum
solutions corresponding to $\omega =\sqrt{\left\vert g\right\vert }=\tau .$
As pointed out in \cite{harvey}, the cases with $a_{1}+1=0$ were ignored by
Kasner$.$ These missed Kasner solutions, were also not considered in the
formalism of \cite{landau}, and belong to a class of solutions with $\omega
=1$ \cite{Parno}.

Melvin solutions \cite{melvin} with a dilaton field were considered in \cite
{GibMae}. Moreover, starting with the four-dimensional scalar-Kasner metric 
\cite{Belinski} and using generating techniques \cite{har, earn}, dilaton
Melvin cosmologies and flux tubes were obtained in \cite{dowker, kt}.

Generalizations of Kasner metric to theories with form gauge fields and a
dilaton field to all space-time dimensions and signatures were considered in 
\cite{s1}. Also generalized Melvin metrics were found for the theories of $
N=2$ supergravity coupled to vector multiplets in four and five dimensions 
\cite{s2, s3}. More recently, non-diagonal solutions for $D$-dimensional
Einstein-Maxwell theory were obtained in \cite{new} through an explicit
analysis of the equations of motion.

In this paper, we are interested in finding general solutions depending on
one variable for the $D$-dimensional theories described by the action 
\begin{equation}
S=\int d^{D}x\sqrt{\left\vert g\right\vert }\left( R-\frac{1}{2}\partial
_{\mu }\phi \partial ^{\mu }\phi -\frac{\varepsilon }{2\,m!}\,e^{\beta \phi
}\,\mathcal{F}_{m}^{2}\right) \text{ }.  \label{ch}
\end{equation}%
The non-trivial fields of the action (\ref{ch}) are the graviton, an $m$
-form field strength $\mathcal{F}_{m}$ and a dilaton field $\phi $. The
constant $\beta $ depends on the details of the theory under consideration
and $\varepsilon =\pm 1$. For a given choice of space-time dimension, rank
of the form gauge field $m$ and dilaton coupling $\beta ,$ the action (\ref
{ch}) describe the bosonic sector of various supergravity theories. Such
theories can be obtained as a truncation of the low energy limit of M-theory
and string theories.

Flux $p$-branes and S-branes solutions to (\ref{ch}) and for $\varepsilon =1$
were analyzed in \cite{gs} and \cite{gal}. The cases with $\varepsilon =-1$
correspond to a phantom form fields. Such fields appear naturally in
supergravity theories descending from M-theory \cite{Hull, liu}. Black hole
solutions for (\ref{ch}) with phantom fields were considered in the
literature (see for example \cite{gm, emd}). Moreover phantom energy was
implemented in the study of cosmology (see for example \cite{cosmo}).

We organize our work as follows. In the next section, we obtain
generalizations of the four-dimensional scalar-Kasner solutions \cite
{Belinski}. In section three, we find solutions for (\ref{ch}) where all the
fields are non-trivial and depend only on one variable. We analyze the
equations of motion derived from (\ref{ch}) and obtain families of solutions
with nontrivial gauge and dilaton fields. Section four contains a summary
and a discussion of our results.

\section{SOLUTIONS WITH\ SCALAR FIELD}
In this section we shall construct solutions to (\ref{ch}) with non-trivial
scalar field. In this case, the equations of motion derived from (\ref{ch})
are simply given by 
\begin{equation}
R_{\mu \nu }=\frac{1}{2}\partial _{\mu }\phi \partial _{\nu }\phi \text{ },
\end{equation}%
and%
\begin{equation}
\partial _{\mu }\left( \omega \partial ^{\mu }\phi \right) =0\text{ }.
\label{sin}
\end{equation}%
As in \cite{landau}, we consider metric solutions of the form 
\begin{equation}
ds^{2}=\epsilon _{0}d\tau ^{2}+g_{\mu \nu }(\tau )dx^{\mu }dx^{\nu },
\end{equation}%
with $\mu ,$ $\nu =1,...,D-1,$ then the analysis of Einstein equations of
motion gives 
\begin{align}
\frac{\ddot{\omega}}{\omega }-\frac{\dot{\omega}^{2}}{\omega ^{2}}+\frac{1}{4
}\kappa ^{\mu }{}_{\nu }\kappa ^{\nu }{}_{\mu }+\frac{1}{2}\dot{\phi}^{2}& =0
\text{ },  \label{h1} \\
\dot{\kappa}^{\mu }{}_{\nu }+\frac{\dot{\omega}}{\omega }\kappa ^{\mu
}{}_{\nu }& =0\text{ },  \label{h2}
\end{align}%
where $\kappa ^{\mu }{}_{\nu }=g^{\mu \rho }\dot{g}_{\rho \nu }$\ \cite
{landau}. The derivative with respect to the variable $\tau $ is denoted by
an overdot$.$ We also obtain from the scalar equation of motion (\ref{sin})
\begin{equation}
\dot{\phi}=\frac{s}{\omega }  \label{ait}
\end{equation}%
where $s$ is a constant. The trace of (\ref{h2}) gives the condition 
\begin{equation}
\ddot{\omega}=0\text{ }.
\end{equation}%
As in \cite{landau}, we first consider the solutions with $\omega =\tau $,
then (\ref{h2}) gives 
\begin{equation}
\kappa ^{\mu }{}_{\nu }=\frac{2}{\tau }\lambda ^{\mu }{}_{\nu }  \label{so}
\end{equation}%
where $\lambda ^{\mu }{}_{\nu }$ are constants satisfying the condition 
\begin{equation}
\lambda ^{\mu }{}_{\mu }=1\text{ }.  \label{f}
\end{equation}%
Plugging the solution (\ref{so}) into (\ref{h1}) gives the condition 
\begin{equation}
\lambda ^{\mu }{}_{\nu }\lambda ^{\nu }{}_{\mu }=1-\frac{1}{2}s^{2}\text{ }.
\label{sq}
\end{equation}
We also deduce from (\ref{h2}) that the metric components satisfy
\begin{equation}
\dot{g}_{\mu \nu }=\frac{2}{\tau }g_{\mu \rho }\lambda ^{\rho }{}_{\nu } 
\text{ }.  \label{do}
\end{equation}
General solutions to ({\ref{do}}) are given by 
\begin{equation}
g_{\mu \nu }=h_{\mu \rho }\left( e^{\left( 2\ln \tau \right) \lambda
}\right) ^{\rho }{}_{\nu }  \label{gensol1}
\end{equation}%
where $h_{\mu \nu }$ are symmetric constants satisfying the condition 
\begin{equation}
h_{\mu \rho }\lambda ^{\rho }{}_{\nu }=h_{\nu \rho }\lambda ^{\rho }{}_{\mu
} \text{ }.  \label{sym2}
\end{equation}

For a given matrix defined by $\lambda ^{\mu }{}_{\nu }$ and satisfying (\ref
{f}) and (\ref{sq})$,$ explicit metric solutions can be constructed using (
\ref{gensol1}) and (\ref{sym2}). The solutions depend on the eigenvalues of
the matrix $\lambda ^{\mu }{}_{\nu }$ and their multiplicity. For diagonal $
\lambda ^{\mu }{}_{\nu }$, we obtain $D$-dimensional generalizations of the
four-dimensional space-time solutions found in \cite{Belinski}.

Solutions for $\omega =1$ should also be considered$.$ As already mentioned,
vacuum solutions corresponding to these cases were ignored in \cite{kasner,
landau}. For the theories with scalar field, we obtain 
\begin{equation}
\kappa ^{\mu }{}_{\nu }=\text{\ }\theta ^{\mu }{}_{\nu }\text{ },\text{ \ \
\ \ }\dot{\phi}=s\text{ ,}
\end{equation}%
where $s$ is a constant and $\lambda ^{\mu }{}_{\nu }$ are constants
satisfying 
\begin{equation}
\theta ^{\mu }{}_{\mu }=0\text{ },\text{ \ \ \ \ }\theta ^{\mu }{}_{\nu
}\theta ^{\nu }{}_{\mu }=-2s^{2}\text{ }.
\end{equation}%
\ The $D$-dimensional metric for this class of solutions takes the form 
\begin{equation}
ds^{2}=\epsilon _{0}d\tau ^{2}+h_{\mu \rho }\left( e^{\theta \tau }\right)
^{\rho }{}_{\nu }dx^{\mu }dx^{\nu },  \label{nw}
\end{equation}%
and the constants $h_{\mu \nu }$ are symmetric and satisfying $h_{\mu \rho
}\theta ^{\rho }{}_{\nu }=h_{\nu \rho }\theta ^{\rho }{}_{\mu }$.

\section{GENERAL\ SOLUTIONS}
In this section, we shall find exact solutions for the theories described by
(\ref{ch}) where all the fields are non-trivial and depend only on one
variable. The gravitational, gauge and scalar field equations derived from (
\ref{ch}) are 
\begin{equation}
R_{\mu \nu }=\frac{1}{2}\partial _{\mu }\phi \partial _{\nu }\phi
+\varepsilon e^{\beta \phi }\left[ \frac{1}{2(m-1)!}\mathcal{F}_{\mu \alpha
_{2}\cdots \alpha _{m}}\mathcal{F}_{\nu }{}^{\alpha _{2}\cdots \alpha _{m}}-
\frac{\left( m-1\right) }{2m!(d-2)}\mathcal{F}_{m}^{2}\,g_{\mu \nu }\right] 
\text{ },  \label{ein}
\end{equation}
and 
\begin{equation}
\partial _{\mu }\left( \omega e^{\beta \phi }\,\mathcal{F}^{\mu \nu
_{2}\cdots \nu _{m}}\right) =0\text{ },  \label{max}
\end{equation}
and$\,$ 
\begin{equation}
\partial _{\mu }\left( \omega \partial ^{\mu }\phi \right) =\frac{\beta
\varepsilon }{2\,m!}\omega e^{\beta \phi }\mathcal{F}_{m}^{2}\text{ }.
\label{seh}
\end{equation}
The first class of solutions we consider is for a $\left( q+1\right) $-form
whose only non-trivial component is given by $\mathcal{F}^{\tau 1...q}.$ Our
metric solution is of the form 
\begin{equation}
ds^{2}=\epsilon _{0}d\tau ^{2}+g_{\mu \nu }(\tau )dx^{\mu }dx^{\nu
}=\epsilon _{0}d\tau ^{2}+\sum_{i=1}^{q}g_{ii}(\tau )\left( dx^{i}\right)
^{2}+g_{ab}(\tau )dx^{a}dx^{b}\text{ },  \label{met}
\end{equation}
where $D=p+q+1.$ For this class of solutions, the gauge equations of motion (
\ref{max}), where all the fields depend only on the coordinate $\tau ,$ can
be solved by 
\begin{equation}
\mathcal{F}^{\tau 1...q}=\frac{Q}{\omega }e^{-\beta \phi }\text{ },
\label{gs}
\end{equation}
where $Q$ is a constant.

The Einstein equations of motion (\ref{ein}) together with (\ref{gs}) give 
\begin{align}
\ddot{\omega}\omega -\dot{\omega}^{2}+\frac{1}{4}\omega ^{2}\kappa ^{\mu
}{}_{\nu }\kappa ^{\nu }{}_{\mu }+\frac{1}{2}\omega ^{2}\dot{\phi}^{2}+\frac{
\varepsilon }{2}\left( \frac{p-1}{D-2}\right) Q^{2}e^{-\beta \phi
}\,\prod_{i=1}^{q}g_{ii}& =0\text{ },  \label{du1} \\
\dot{\kappa}^{i}{}_{i}+\frac{\dot{\omega}}{\omega }\kappa
^{i}{}_{i}+\varepsilon \left( \frac{p-1}{D-2}\right) \frac{Q^{2}}{\omega ^{2}
}e^{-\beta \phi }\,\prod_{i=1}^{q}g_{ii}& =0\text{ },  \label{du2} \\
\dot{\kappa}^{a}{}_{b}+\frac{\dot{\omega}}{\omega }\kappa
^{a}{}_{b}-\varepsilon \frac{q}{(D-2)}\frac{Q^{2}}{\omega ^{2}}e^{-\beta
\phi }\delta ^{a}{}_{b}\prod_{i=1}^{q}g_{ii}& =0\text{ }.  \label{du3}
\end{align}
In (\ref{du2}) and in what follows we will not adopt summation convention
for repeated $i$ indices. Note that the relations (\ref{du2}) and (\ref{du3}
) imply the condition 
\begin{equation}
\ddot{\omega}=\frac{\varepsilon qQ^{2}}{2\left( D-2\right) \omega }
\,e^{-\beta \phi }\prod_{i=1}^{q}g_{ii}\text{ }.  \label{tr2}
\end{equation}
Using (\ref{tr2}), then (\ref{du1}), (\ref{du2}) and (\ref{du3}) reduce to 
\begin{align}
2\frac{\left( D-2\right) }{q}\,\frac{\ddot{\omega}}{\omega }-2\frac{\dot{%
\omega}^{2}}{\omega ^{2}}+\frac{1}{2}\kappa ^{\mu }{}_{\nu }\kappa ^{\nu
}{}_{\mu }+\dot{\phi}^{2}& =0\text{ },  \label{j1} \\
\dot{\kappa}^{i}{}_{i}+\frac{\dot{\omega}}{\omega }\kappa ^{i}{}_{i}+2\frac{
\left( p-1\right) }{q}\frac{\ddot{\omega}}{\omega }& =0\text{ },  \label{jj}
\\
\dot{\kappa}^{a}{}_{b}+\frac{\dot{\omega}}{\omega }\kappa ^{a}{}_{b}-2\frac{
\ddot{\omega}}{\omega }\delta ^{a}{}_{b}& =0\text{ }.  \label{jjj}
\end{align}%
The scalar equation of motion (\ref{seh}) gives 
\begin{equation}
\dot{\omega}\dot{\phi}+\omega \ddot{\phi}=\beta \left( \frac{D-2}{q}\right) 
\ddot{\omega}\text{ }.  \label{sc}
\end{equation}%
Eqs. (\ref{jj}) and (\ref{jjj}) can be integrated and give 
\begin{eqnarray}
\kappa ^{i}{}_{i} &=&\frac{1}{\omega }\left[ \theta ^{i}{}_{i}-2\left( \frac{
p-1}{q}\right) \dot{\omega}\right] \text{ },  \notag \\
\kappa ^{a}{}_{b} &=&\frac{1}{\omega }\left( \theta ^{a}{}_{b}+2\dot{\omega}
\delta ^{a}{}_{b}\right) \text{ },  \label{in1}
\end{eqnarray}
with the condition 
\begin{equation}
\theta ^{\mu }{}_{\mu }=\theta ^{a}{}_{a}+\sum_{i=1}^{q}\theta ^{i}{}_{i}=0
\text{ . }  \label{tr}
\end{equation}%
Substituting (\ref{in1}) into the relation (\ref{j1}) we obtain 
\begin{equation}
\frac{1}{2}\theta ^{\mu }{}_{\nu }\theta ^{\nu }{}_{\mu }+2\left( \frac{D-2}{
q}\right) \left[ \left( p-1\right) \dot{\omega}^{2}-\dot{\omega}
\sum_{i=1}^{q}\theta ^{i}{}_{i}\right] +\varepsilon Q^{2}e^{-\beta \phi
}\,\prod_{i=1}^{q}g_{ii}+\omega ^{2}\dot{\phi}^{2}=0\text{ }.  \label{cs}
\end{equation}
Moreover, Eqs. (\ref{in1}) imply that the space-time metric components
satisfy 
\begin{align}
\dot{g}_{ii}& =\frac{1}{\omega }\left( \theta ^{i}{}_{i}g_{ii}-2\left( \frac{
p-1}{q}\right) \dot{\omega}g_{ii}\right) \text{ },  \label{me1} \\
\dot{g}_{ab}& =\frac{1}{\omega }\left( g_{ac}\theta ^{c}{}_{b}+2\dot{\omega}
g_{ab}\right) \text{ }.  \label{me2}
\end{align}
To obtain some explicit solutions, we perform the convenient change of
variables \cite{new} 
\begin{equation}
\frac{d\sigma }{d\tau }=\frac{1}{H}\text{ },\text{ \ \ \ \ \ \ \ \ }\omega
=\sigma H\text{ },  \label{cv}
\end{equation}%
and define the new parameters 
\begin{equation}
\lambda ^{i}{}_{i}=\frac{1}{2}\left( \theta ^{i}{}_{i}-2\frac{\left(
p-1\right) }{q}\right) \text{ },\text{ \ \ }\lambda _{\text{ }b}^{a}=\frac{1
}{2}\left( \theta ^{a}{}_{b}+2\delta ^{a}{}_{b}\right) \text{ }.  \label{cov}
\end{equation}
Eqs. (\ref{me1}) and (\ref{me2}) can then be solved by 
\begin{align}
g_{ii}& =h_{ii}\sigma ^{2\lambda ^{i}{}_{i}}H^{-2\left( \frac{p-1}{q}\right)
}\text{ },  \notag \\
g_{ab}& =h_{ac}\left( e^{2\lambda \log \sigma }\right) ^{c}{}_{b}H^{2},
\end{align}%
with $h_{ii}=\pm 1$ and the symmetric constants $h_{\mu \nu }$ satisfying $
h_{\mu \rho }\lambda ^{\rho }{}_{\nu }=h_{\nu \rho }\lambda ^{\rho }{}_{\mu
} $.

The scalar equation of motion (\ref{sc}), in terms of the new variables,
gives 
\begin{equation}
\frac{d}{d\sigma }\left( \sigma \frac{d\phi }{d\sigma }\right) =\beta \left( 
\frac{D-2}{q}\right) \frac{d}{d\sigma }\left( \sigma \frac{d}{d\sigma }\log
H\right)
\end{equation}%
and admits the solutions (ignoring a multiplicative constant) 
\begin{equation}
e^{\phi }=\sigma ^{s}H^{\beta \left( \frac{D-2}{q}\right) }  \label{ss}
\end{equation}%
where $s$ is a constant. The function $H$ can be determined using Eq. (\ref
{tr2}) which in terms of the new variables reads 
\begin{equation}
\frac{1}{H}\frac{dH}{d\sigma }-\frac{\sigma }{H^{2}}\left( \frac{dH}{d\sigma 
}\right) ^{2}+\frac{\sigma }{H}\frac{d^{2}H}{d\sigma ^{2}}=\frac{\varepsilon
q}{2\left( D-2\right) }\prod_{i=1}^{q}h_{ii}\sigma ^{\left( -\beta
s+2l-1\right) }Q^{2}H^{2-2p-\beta ^{2}\left( \frac{D-2}{q}\right) }\text{ }.
\label{he}
\end{equation}
where we have defined 
\begin{equation}
l=\sum_{i=1}^{q}\lambda ^{i}{}_{i}\text{ }.
\end{equation}
Eq. (\ref{he}) admits the solution 
\begin{equation}
H=\left( 1+c\sigma ^{-\beta s+2l}\right) ^{\gamma }\text{ },
\end{equation}
with 
\begin{eqnarray}
\frac{1}{\gamma } &=&p-1+\beta ^{2}\left( \frac{D-2}{2q}\right) \text{ }, \\
c &=&\frac{\varepsilon qQ^{2}\left[ p-1+\beta ^{2}\left( \frac{D-2}{2q}
\right) \right] }{2\left( D-2\right) \left( 2l-\beta s\right) ^{2}}%
\prod_{i=1}^{q}h_{ii}\text{ }.
\end{eqnarray}
Using (\ref{cov}) then (\ref{tr}) becomes 
\begin{equation}
\lambda ^{\mu }{}_{\mu }=\sum_{i=1}^{q}\lambda ^{i}{}_{i}+\lambda
^{a}{}_{a}=1\text{ }.  \label{c1}
\end{equation}
Eq. (\ref{cs}), for our solutions, reduces to the condition 
\begin{equation}
\frac{1}{2}\theta ^{\mu }{}_{\nu }\theta ^{\nu }{}_{\mu }+2\left( \frac{D-2}{
q}\right) \left( p-1+\theta ^{a}{}_{a}\right) +s^{2}=0\text{ }.
\end{equation}
Note that with the change of constants as given by (\ref{cov}), this
condition simply reduces to 
\begin{equation}
\sum_{i=1}^{q}\left( \lambda ^{i}{}_{i}\right) ^{2}+\lambda _{\text{ }
a}^{b}\lambda _{\text{ }b}^{a}=1-\frac{s^{2}}{2}\text{ }.  \label{c2}
\end{equation}
To summarize, our solutions are given by 
\begin{eqnarray}
ds^{2} &=&H^{2}\left( \epsilon _{0}d\sigma ^{2}+h_{ac}\left( e^{2\lambda
\log \sigma }\right) ^{c}{}_{b}dx^{a}dx^{b}\right)
+\sum_{i=1}^{q}h_{ii}\sigma ^{2\lambda ^{i}{}_{i}}H^{-2\left( \frac{p-1}{q}
\right) }\left( dx^{i}\right) ^{2}\text{ },  \notag \\
\mathcal{F}^{\sigma 1...q} &=&\frac{Q}{\sigma ^{\beta s+1}}H^{-\beta
^{2}\left( \frac{D-2}{q}\right) -2}\text{ },  \notag \\
e^{\phi } &=&\sigma ^{s}H^{\beta \left( \frac{D-2}{q}\right) }\text{ },
\end{eqnarray}
with the conditions (\ref{c1}), (\ref{c2}) and 
\begin{equation}
H^{\left( p-1+\beta ^{2}\left( \frac{D-2}{2q}\right) \right) }=\left(
1+\varepsilon \prod_{i=1}^{q}h_{ii}\frac{q\left[ p-1+\beta ^{2}\left( \frac{
D-2}{2q}\right) \right] }{2(D-2)\left( 2l-\beta s\right) ^{2}}Q^{2}\text{ }
\sigma ^{2l-\beta s}\right) \text{ }.
\end{equation}
Next we consider another class of solutions for which we have a non-trivial $
p$-form field given by 
\begin{equation}
\mathcal{F}_{12...p}=P\text{ },  \label{mag}
\end{equation}%
where $P$ is a constant, $D=p+q+1$ and our metric is of the form
\begin{equation}
ds^{2}=\epsilon _{0}d\tau ^{2}+g_{\mu \nu }(\tau )dx^{\mu }dx^{\nu
}=\epsilon _{0}d\tau ^{2}+\sum_{i=1}^{p}g_{ii}(\tau )\left( dx^{i}\right)
^{2}+g_{ab}(\tau )dx^{a}dx^{b}\text{ }.  \label{me}
\end{equation}%
The analysis of the Einstein gravitational equations of motion (\ref{ein})
then gives 
\begin{align}
\frac{\left( D-2\right) }{q}\frac{\ddot{\omega}}{\omega }-\frac{\dot{\omega}
^{2}}{\omega ^{2}}+\frac{1}{4}\kappa ^{\mu }{}_{\nu }\kappa ^{\nu }{}_{\mu }+
\frac{1}{2}\dot{\phi}^{2}& =0\text{ },  \label{qw1} \\
\dot{\kappa}^{i}{}_{i}+\frac{\dot{\omega}}{\omega }\kappa ^{i}{}_{i}-2\frac{
\ddot{\omega}}{\omega }& =0\text{ },  \label{qw2} \\
\dot{\kappa}^{a}{}_{b}+\frac{\dot{\omega}}{\omega }\kappa ^{a}{}_{b}+2\frac{
\left( p-1\right) }{q}\frac{\ddot{\omega}}{\omega }\delta ^{a}{}_{b}& =0%
\text{ }.  \label{qw3}
\end{align}
As well as the relation 
\begin{equation}
\ddot{\omega}=-\varepsilon \epsilon _{0}\frac{q\omega e^{\beta \phi }P^{2}}{
2(D-2)}\prod_{i=1}^{p}g^{ii}\text{ }.  \label{tt}
\end{equation}%
Eqs. (\ref{qw2}) and (\ref{qw3}) can be integrated to give the relations 
\begin{align}
\kappa ^{i}{}_{i}& =\frac{1}{\omega }\left( \theta ^{i}{}_{i}+2\dot{\omega}
\right) \text{ },  \label{k1} \\
\kappa ^{a}{}_{b}& =\frac{1}{\omega }\left[ \theta ^{a}{}_{b}-\frac{2(p-1)}{q
}\dot{\omega}\delta ^{a}{}_{b}\right]  \label{k2}
\end{align}
with the condition 
\begin{equation}
\sum_{i=1}^{p}\theta ^{i}{}_{i}+\theta ^{a}{}_{a}=0\text{ }.  \label{se}
\end{equation}%
Upon substituting (\ref{k1}) and (\ref{k2}) into (\ref{qw1}) we obtain 
\begin{equation}
\frac{1}{2}\theta ^{\mu }{}_{\nu }\theta ^{\nu }{}_{\mu }+2\left( \frac{D-2}{
q}\right) \left[ (p-1)\dot{\omega}^{2}-\dot{\omega}\theta ^{a}{}_{a}\right]
+\omega ^{2}\left( \dot{\phi}^{2}-\varepsilon \epsilon _{0}e^{\beta \phi
}P^{2}\prod_{i=1}^{p}g^{ii}\right) =0\text{ }.  \label{check}
\end{equation}
These relations also give for the metric components 
\begin{align}
\dot{g}_{ii}& =\frac{1}{\omega }\left( \theta ^{i}{}_{i}+2\dot{\omega}
\right) g_{ii}\text{ },  \notag \\
\dot{g}_{ab}& =\frac{1}{\omega }\left( g_{ac}\theta ^{c}{}_{b}-\frac{2(p-1)}{
q}\dot{\omega}g_{ab}\right) \text{ }.  \label{met2}
\end{align}
The scalar equation of motion reads 
\begin{equation}
\,\partial _{\tau }\left( \omega \dot{\phi}\right) =-\beta \frac{(D-2)}{q}
\ddot{\omega}\text{ }.  \label{seom}
\end{equation}
To get some explicit solutions, we perform again the change of variables
given in (\ref{cv}) and obtain from (\ref{met2}) the solutions 
\begin{align*}
g_{ii}& =h_{ii}\sigma ^{2\lambda ^{i}{}_{i}}H^{2}, \\
g_{ab}& =h_{ac}\left( e^{2\lambda \log \sigma }\right) ^{c}{}_{b}H^{-2\frac{
(p-1)}{q}}.
\end{align*}
where we have defined 
\begin{equation}
\lambda ^{i}{}_{i}=\frac{1}{2}\left( \theta ^{i}{}_{i}+2\right) ,\ \ \
\lambda ^{a}{}_{b}=\frac{1}{2}\left[ \theta ^{a}{}_{b}-2\frac{(p-1)}{q}
\delta ^{a}{}_{b}\right] \text{ }.  \label{fa}
\end{equation}
Again $h_{ii}=\pm 1$ and $h_{ab}$ is symmetric and satisfies $h_{\mu \rho
}\lambda ^{\rho }{}_{\nu }=h_{\nu \rho }\lambda ^{\rho }{}_{\mu }$.

The scalar equation of motion (\ref{seom}), in terms of the new variables,
reduces to the simple equation 
\begin{equation}
\frac{d}{d\sigma }\left( \sigma \frac{d\phi }{d\sigma }\right) =-\beta
\left( \frac{D-2}{q}\right) \frac{d}{d\sigma }\left[ \sigma \left( \frac{d}{
d\sigma }\left( \log H\right) \right) \right]
\end{equation}
which can be easily seen to admit the solution 
\begin{equation}
e^{\phi }=\sigma ^{s}H^{-\beta \left( \frac{D-2}{q}\right) }\text{ }.
\end{equation}
In terms of the new variables, Eq. (\ref{tt}) gives 
\begin{equation}
\frac{1}{H^{2}}\left( \frac{dH}{d\sigma }-\frac{\sigma }{H}\left( \frac{dH}{
d\sigma }\right) ^{2}+\sigma \frac{d^{2}H}{d\sigma ^{2}}\right) =-\frac{
\epsilon _{0}\varepsilon q}{2\left( D-2\right) }\prod_{i=1}^{p}h_{ii}P^{2}
\sigma ^{\left( 2r+\beta s-1\right) }H^{1-2p-\beta ^{2}\left( \frac{D-2}{q}
\right) }\text{ }.
\end{equation}
where we have defined 
\begin{equation}
r=1-\sum_{i=1}^{p}\lambda ^{i}{}_{i}\text{ }.
\end{equation}%
This admits the solution 
\begin{equation}
H=\left( 1+c\sigma ^{2r+\beta s}\right) ^{\gamma }
\end{equation}%
with%
\begin{eqnarray}
\frac{1}{\gamma } &=&p-1+\beta ^{2}\left( \frac{D-2}{2q}\right) \text{ }, \\
c &=&-\epsilon _{0}\varepsilon \prod_{i=1}^{p}h_{ii}\frac{q\left[ p-1+\beta
^{2}\left( \frac{D-2}{2q}\right) \right] }{2\left( D-2\right) \left(
2r+\beta s\right) ^{2}}P^{2}\text{ .}
\end{eqnarray}%
We now turn back to Eq. (\ref{check}) from which we obtain the condition 
\begin{equation}
\frac{1}{2}\theta ^{\mu }{}_{\nu }\theta ^{\nu }{}_{\mu }+2\left( \frac{D-2}{
q}\right) \left( p-1-\theta ^{a}{}_{a}\right) +s^{2}=0\text{ }.  \label{gra}
\end{equation}%
Using (\ref{fa}), the conditions (\ref{se}) and (\ref{gra}) give 
\begin{eqnarray}
\lambda ^{\mu }{}_{\mu } &=&1\text{ },  \label{mf} \\
\lambda ^{\mu }{}_{\nu }\lambda ^{\nu }{}_{\mu } &=&1-\frac{s^{2}}{2}\text{ }
.  \label{mff}
\end{eqnarray}%
To summarize, our second class of solutions for which $\mathcal{F}
_{1....p}=P $ are given by 
\begin{eqnarray}
ds^{2} &=&H^{2}\left( \epsilon _{0}d\sigma ^{2}+\sum_{i=1}^{p}h_{ii}\sigma
^{2\lambda ^{i}{}_{i}}\left( dx^{i}\right) ^{2}\right) +h_{ac}\left(
e^{2\lambda \log \sigma }\right) ^{c}{}_{b}H^{-2\frac{(p-1)}{q}}dx^{a}dx^{b}
\text{ },  \notag \\
e^{\phi } &=&\sigma ^{s}H^{-\beta \left( \frac{D-2}{q}\right) }\text{ }, 
\notag \\
H^{\left[ p-1+\beta ^{2}\left( \frac{D-2}{2q}\right) \right] } &=&1-\epsilon
_{0}\varepsilon \prod_{i=1}^{p}h_{ii}\frac{q\left[ p-1+\beta ^{2}\left( 
\frac{D-2}{2q}\right) \right] }{2\left( D-2\right) \left( 2r+\beta s\right)
^{2}}P^{2}\sigma ^{2r+\beta s}
\end{eqnarray}
supplemented with the conditions (\ref{mf}) and (\ref{mff}).

\bigskip More classes of solutions can be obtained if we consider the
following change of variable 
\begin{equation}
\frac{d\chi }{d\tau }=\frac{1}{\omega (\chi )}\text{ }.
\end{equation}
Starting with solutions of the form (\ref{met}) and where the gauge field is
given by (\ref{gs}). In terms of the new coordinates, the Eqs. (\ref{me1})
and (\ref{me2}) defining the metric components become 
\begin{align}
g_{ii}^{\prime }& =\theta ^{i}{}_{i}g_{ii}-2\left( \frac{p-1}{q}\right) 
\frac{\omega ^{\prime }}{\omega }g_{ii}\text{ }, \\
g_{ab}^{\prime }& =g_{ac}\theta ^{c}{}_{b}+2\frac{\omega ^{\prime }}{\omega }
g_{ab}\text{ },
\end{align}
where the prime denotes differentiation with respect to $\chi .$ These Eqs.
admit the solutions 
\begin{align}
g_{ii}& =h_{ii}e^{\chi \theta ^{i}{}_{i}}\omega ^{-2\left( \frac{p-1}{q}
\right) }\text{ }, \\
g_{ab}& =h_{ac}\left( e^{\theta \chi }\right) ^{c}{}_{b}\omega ^{2}\text{ }.
\end{align}
with 
\begin{equation}
\theta ^{\mu }{}_{\mu }=\sum_{i=1}^{q}\theta ^{i}{}_{i}+\theta ^{a}{}_{a}=0
\text{ }.  \label{dan}
\end{equation}
The scalar equation of motion (\ref{sc}) gives 
\begin{equation}
\phi ^{\prime \prime }=\beta \left( \frac{D-2}{q}\right) \left( \log \omega
\right) ^{\prime \prime }
\end{equation}
and admits the solution 
\begin{equation}
e^{\phi }=e^{s\chi }\omega ^{\beta \left( \frac{D-2}{q}\right) }\text{ }.
\end{equation}
with constant $s.$ The function $\omega $ can be determined via the relation
(\ref{tr2}) which in terms of new variables reduces to%
\begin{equation}
\left( \log \omega \right) ^{\prime \prime }-e^{-\chi \left( \beta s+\theta
^{a}{}_{a}\right) }\omega ^{2(1-p)-\beta ^{2}\left( \frac{D-2}{q}\right) }%
\frac{\varepsilon qQ^{2}}{2\left( D-2\right) }\prod_{i=1}^{q}h_{ii}=0
\end{equation}
and can be solved by 
\begin{equation}
\omega =\left( 1+\mu e^{\left( 2l-\beta s\right) \chi }\right) ^{\gamma }
\end{equation}
with 
\begin{eqnarray}
\frac{1}{\gamma } &=&p-1+\beta ^{2}\left( \frac{D-2}{2q}\right) \text{ }, 
\notag \\
2l &=&-\theta ^{a}{}_{a}\text{ },  \notag \\
\mu &=&\varepsilon \frac{q\left[ p-1+\beta ^{2}\left( \frac{D-2}{2q}\right) %
\right] }{2\left( D-2\right) \left( 2l-\beta s\right) ^{2}}
Q^{2}\,\prod_{i=1}^{q}h_{ii}\text{ }.
\end{eqnarray}
Upon substituting all in (\ref{cs}), we obtain \bigskip the simple condition 
\begin{equation}
\theta ^{\mu }{}_{\nu }\theta ^{\nu }{}_{\mu }=\sum_{i=1}^{q}\left( \theta
^{i}{}_{i}\right) ^{2}+\theta ^{a}{}_{b}\theta ^{b}{}_{a}\text{ }=-2s^{2}
\text{ }.  \label{dam}
\end{equation}
For these solutions we have 
\begin{equation}
\mathcal{F}^{\chi 1...q}=\frac{Q}{\omega ^{2+\beta ^{2}\left( \frac{D-2}{q}
\right) }}e^{-\beta s\chi }\text{ }.
\end{equation}
Performing similar analysis for solutions where the space-time metric and $p$
-form field strength are given by (\ref{me}) and (\ref{mag}), a new family
of solutions can be obtained and is given by
\begin{eqnarray}
ds^{2} &=&\omega ^{2}\left( \epsilon _{0}d\chi
^{2}+\sum_{i=1}^{p}h_{ii}e^{\chi \theta ^{i}{}_{i}}\left( dx^{i}\right)
^{2}\right) +h_{ac}\left( e^{\theta \chi }\right) ^{c}{}_{b}\omega ^{-2\frac{
(p-1)}{q}}dx^{a}dx^{b}\text{ },  \notag \\
\omega ^{p-1+\beta ^{2}\left( \frac{D-2}{2q}\right) } &=&1-\varepsilon
\epsilon _{0}\frac{q\left[ \left( p-1+\beta ^{2}\left( \frac{D-2}{2q}\right)
\right) \right] }{2\left( \theta ^{a}{}_{a}+\beta s\right) ^{2}\left(
D-2\right) }\text{ }P^{2}e^{\left( \beta s+\theta ^{a}{}_{a}\right) \chi
}\prod_{i=1}^{p}h_{ii}\text{ },  \notag \\
\mathcal{F}_{1...p} &=&P\text{ },  \notag \\
e^{\phi } &=&e^{s\chi }\omega ^{-\beta \left( \frac{D-2}{q}\right) }\text{ },
\end{eqnarray}
with the conditions
\begin{eqnarray}
\theta ^{\mu }{}_{\mu } &=&\sum_{i=1}^{p}\theta ^{i}{}_{i}+\theta
^{a}{}_{a}=0\text{ },  \notag \\
\theta ^{\mu }{}_{\nu }\theta ^{\nu }{}_{\mu } &=&\sum_{i=1}^{p}\left(
\theta ^{i}{}_{i}\right) ^{2}+\theta ^{a}{}_{b}\theta ^{b}{}_{a}\text{ }
=-2s^{2}\text{ }.
\end{eqnarray}
For the last two families of solutions we also have $h_{ii}=\pm 1$ and the
condition that $h_{\mu \rho }$ is symmetric and satisfies 
\begin{equation}
h_{\mu \rho }\theta ^{\rho }{}_{\nu }=h_{\nu \rho }\theta ^{\rho }{}_{\mu }
\text{ }.
\end{equation}

\section{SUMMARY AND\ DISCUSSION}
In this paper, we obtained families of general solutions depending on one
variable for gravitational theories with a dilaton and an $m$-form gauge
field strength. Our solutions are valid for arbitrary space-time dimensions
and signatures. In ten space-time dimensions, our general Lagrangian (\ref
{ch}), for a choice of gauge fields and dilaton coupling, represent the
bosonic part of many ten-dimensional supergravity theories. These theories
include the standard type IIA and type IIB supergravity as well as the
exotic type IIA$^{\ast }$, \ type IIB$^{\ast }$ and type IIB$^{\prime }$
supergravity theories constructed by Hull \cite{Hull}. Moreover, switching
off the dilaton field, our solutions are also relevant to eleven-dimensional
supergravity theories with 4-form $\mathcal{F}_{4}$ with $(t,s)$ space-time
signatures, $t$ being the number of the time directions and $s$ the number
of spatial ones. We have $\varepsilon =1$ for signatures $\left( 1,10\right) 
$ , $\left( 5,6\right) $ and $\left( 9,2\right) .$ For the mirror theories
with signatures $\left( 10,1\right) $, $\left( 6,5\right) $ and $\left(
2,9\right) $ we have $\varepsilon =-1$ \cite{Hull}$.$ It should be noted
that one should be able to obtain our solutions from vacuum solutions via
solution generating techniques \cite{har, earn}. The seed solutions belong
to the families of vacuum solutions with $\omega =\tau $ and $\omega =1.$

To construct explicit solutions of the general solutions presented, one
needs to identify all the Jordan normal forms of the constrained $\lambda
^{\mu }{}_{\nu }$ and $\theta ^{\mu }{}_{\nu }$ matrices appearing in our
solutions. These Jordan forms naturally depend on the eigenvalues of these
matrices and their multiplicities. It is of importance to investigate the
relevance of our solutions to string cosmology and the study of cosmological
singularities \cite{book}.


\bigskip

\bigskip

\begin{thebibliography}{99}
\bibitem{macel} G. F. R. Ellis and M. A. H. MacCallum, A class of
homogeneous cosmological models, Commun. Math. Phys. \textbf{12, }108 (1969).

\bibitem{kasner} E. Kasner, Geometrical theorems on Einstein's cosmological
equations, Am. J. Math. \textbf{43, } 217 (1921).

\bibitem{harvey} A. Harvey, Complex transformation of the Kasner metric,%
\textit{\ }Gen. Relativ. Gravit, \textbf{21} 1021 (1989).

\bibitem{landau} L. D. Landau and E. M. Lifshitz, The Classical Theory of
Fields\textit{, }Butterworth-Heinemann (1980); E. M. Lifshitz and I. M.
Khalatnikov, Investigations in relativistic cosmology, Adv. Phys. \textbf{12,%
} 185 (1963).

\bibitem{Parno} S. Parnovsky, Vacuum solutions of Einstein equations that
depend on one variable, Bulletin of Taras Shevchenko National University of
Kyiv. Astronomy Vol. 62, P. (2020) 10.

\bibitem{melvin} M. A. Melvin, Pure magnetic and electric geons, Phys. Lett. 
\textbf{8, } 65 (1964).

\bibitem{GibMae} G.~W.~Gibbons and K.-i.~Maeda, Black holes and membranes in
higher dimensional theories with dilaton fields, Nucl.\ Phys.\ B \textbf{298}
, (1988) 741.

\bibitem{Belinski} V.~A.~Belinski and I.~M.~Khalatnikov, Effect of scalar
and vector fields on the nature of the cosmological singularity,\ Sov.\
Phys.\ JETP \textbf{36} (1973) 591.

\bibitem{har} B. K. Harrison, New solutions of the Einstein-Maxwell
equations from old, J. Math. Phys. \textbf{9}, (1968) 1744.

\bibitem{earn} F. J. Ernst, Black holes in a magnetic universe, \ J. Math.
Phys. \textbf{17}, 54 (1976).

\bibitem{dowker} F.~Dowker, J.~P.~Gauntlett, D.~A.~Kastor and
J.~H.~Traschen, Pair creation of dilaton black holes,\ Phys.\ Rev.\ D 
\textbf{49}, (1994) 2909.

\bibitem{kt} D. Kastor and J. Traschen, Melvin magnetic fluxtube/cosmology
correspondence\textit{, }Classical Quantum Gravity \textbf{32}, 235027
(2015).

\bibitem{s1} W. A. Sabra, Kasner branes with arbitrary signature, Phys.
Lett. \textbf{B809}, 135694 (2020).

\bibitem{s2} W. A. Sabra, Kasner metrics and very special geometry, Phys.
Lett. \textbf{B833,} 137380 (2022).

\bibitem{s3} W. A. Sabra, Melvin space-times in supergravity, Phys. Lett. 
\textbf{B847,} 138307 (2023).

\bibitem{new} W. A. Sabra, Metrics depending on one variable in
Einstein-Maxwell theory, Phys. Rev. D \textbf{110}, 064080 (2024).

\bibitem{gs} M. Gutperle and A. Strominger, Fluxbranes in string theory%
\textit{, } J. High Energy Phys. 06 (2001) 035.

\bibitem{gal} C-M Chen, D. V. Gal'tsov and M. Gutperle, S brane solutions in
supergravity theories, Phys. Rev. D \textbf{66,} (2002) 024043.

\bibitem{Hull} C. M. Hull, Duality and the signature of space-time\textit{,}
J. High Energy Phys. \textbf{11} (1998) 017.

\bibitem{liu} J. T. Liu, W. A. Sabra and W. Y. Wen, Consistent reductions of
IIB\textit{*}/M\textit{*} and de Sitter supergravity, J. High Energy Phys.
01 (2004) 007.

\bibitem{gm} G.\thinspace W. Gibbons and D.\thinspace A. Rasheed, Dyson
pairs and zero-mass black holes, Nucl. Phys. \textbf{B476,} (1996) 515.

\bibitem{emd} G. Cl\'{e}ment, J. C. Fabris, and M. E. Rodrigues, Phantom
black holes in Einstein-Maxwell-dilaton theory,\textit{\ }Phys. Rev. D 
\textbf{79,} (2009) 064021.

\bibitem{cosmo} R. R. Caldwell, M. Kamionkowski and N. N. Weinberg, Phantom
energy: Dark energy with $\omega <-1$ causes a cosmic doomsday, Phys. Rev.
Lett. \textbf{91,} (2003) 071301.

\bibitem{book} V. Belinski, and Marc Henneaux,The Cosmological Singularity, Cambridge University Press, Cambridge, England.
\end{thebibliography}
\end{document}